\documentclass[aps,pra,twocolumn]{revtex4}

\newcommand {\be}{\begin{equation}}
\newcommand {\ee}{\end{equation}}
\newcommand {\bey}{\begin{eqnarray}}
\newcommand {\eey}{\end{eqnarray}}

\usepackage{epsfig}

\begin{document}

\title{Macroscopic Quantum Coherence in a repulsive Bose-Einstein condensate.}
\author{A.~Montina and F.T.~Arecchi   \\
Dipartimento di Fisica, Universit\`a di Firenze,\\
Istituto Nazionale di Ottica Applicata, Largo E.~Fermi 6,50125 Firenze,
Italy.}
\date{\today}

\begin{abstract}
We consider a Bose-Einstein bicondensate (BEC) of $^{87}Rb$, trapped in two different 
internal levels, in a situation where the  density undergoes a 
symmetry breaking in momentum space. This occurs for a suitable number of condensed 
atoms within a double well dispersion curve, obtained by 
Raman coupling two internal states with two tilted and detuned light fields.
Evidence of bistability results from the Gross-Pitaevskii equation.
By second quantization, we evaluate the tunneling rate between the two asymmetric states;
the effects of losses on coherence are also considered.
\end{abstract}
\maketitle
The transition from a superfluid to a Mott insulator has been recently demonstrated for
a Bose-Einstein condensate (BEC), made of atoms with repulsive mutual interactions in 
a lattice potential~\cite{Hansch}. Such a
phenomenon goes beyond the mean field approximation and its explanation requires to take
into account
the quantum fluctuations of the matter field. When the quantum tunneling between adjacent
sites dominates the interaction energy, the probability distribution for the atomic 
occupation of a single
site is Poissonian. In the opposite case the minimum energy is obtained reducing the
quantum fluctuation of the local occupation number. 

In a previous paper we have studied a
similar problem, but with only two wells and an attractive interaction~\cite{OnlyMine};
In this case the opposite effect occurs, that is, raising the atom number, i.e. the 
interaction energy, the minimum energy state is obtained increasing the atomic fluctuation
in each site. This is indirectly demonstrated by the numerical observation of the 
symmetry breaking at a critical number of atoms. Approaching the threshold
value, the quantum fluctuations increase, whereas above that value they blow up and a new 
minimum quantum state appears. The associated probability distribution of the condensate 
barycenter displays two peaks, that can be considered as the dead and alive states 
of a Schr\"odinger cat (SC), whose coherent superposition is called 
{\it Macroscopic Quantum Coherence} (MQC). This term was introduced to describe the 
coherent superposition of two macroscopically distinct quantum states that differ 
for the value of a collective variable~\cite{leggett}. 
The phenomenon is observable only for attractive interactions, that tend to cluster the
atoms in one of the two wells. By contrast, a repulsive interaction tends to reduce the
quantum fluctuations and to distribute the same number of atoms in each well, as observed
in Ref.~\onlinecite{Hansch}. MQC has been observed with trapped ions~\cite{Monroe} and
microwave fields in high-Q cavity~\cite{Brune}.

In this work, we discuss the
feasibility of MQC in a BEC of mutually repulsive atoms.
A repulsive interaction acts in the momentum space as
an attractive interaction, therefore we expect that for a double well dispersion curve
the symmetry breaking occurs in the reciprocal space. Such a
dispersion curve can be obtained by two detuned and tilted light fields, that
transfer a net momentum to atoms as they jump from an internal state to another. We
study the problem by finding the stationary
solutions of two coupled Gross-Pitaevskii equations (GP) discretized over a space
lattice. The system undergoes symmetry breaking
in momentum space for a suitable number of atoms and two new stationary states are created.
We then introduce a quantum two mode model,
with the two modes chosen in such a way as to reproduce the stationary solutions of GP,
and evaluate the quantum fluctuation-mediated tunneling rate between the two 
asymmetrical states.
If the coupling with the environment is negligible, MQC occurs between these states. 
A Raman scheme for creating a superposition state with two $Rb$ condensates in
different internal quantum levels
has already been discussed in Refs.~\onlinecite{Cirac,Gordon}; however both papers 
limit themselves to co-propagating light beams, and this implies applicability problems, 
as discussed in Ref.~\onlinecite{OnlyMine}.

Here we refer to $^{87}Rb$ atoms, but our numerical results apply also to 
$^{23}Na$, if some parameters are rescaled. In a previous work~\cite{RMO} 
we considered atoms in two different hyperfine levels ($F=1$, $m_F=-1$ and $F=2$, $m_F=1$), 
however the associated depletion rate~\cite{corn} is too high for our purposes. Here we 
consider condensate atoms that are optically trapped in the two Zeeman levels $F=1$,
$m_F=-1$ and $F=1$, $m_F=1$. An all optical condensation has been reported in 
Ref.~\onlinecite{Barrett}, alternatively, the condensate can be created with a magnetic
confinement  and transferred into an optical trap~\cite{opticaltrap}.
A homogeneous magnetic field has to be applied to remove
the energy degeneracy. These levels are quasi-resonantly coupled by means of two Raman 
fields $L$ and $R$.
We call $\psi_0$ and $\psi_1$ the fields associated with the $m_F=-1$ and $m_F=1$ levels, 
respectively. Furthermore we call $\psi_2$ the upper state of the $D_1$ transition.

The starting equations are
\be\label{part1}
i\hbar\dot\psi_{0,1}=(H_{0,1}-\hbar\omega_{0,1})\psi_{0,1}
+\hbar E_{L,R}(t)e^{-i(\vec k_{L,R}\cdot\vec x-\omega_{L,R}t)}\psi_2
\ee
\be\label{part3}
i\hbar\dot\psi_2=\hbar\omega_2\psi_2
+[\hbar E_L^*(t)e^{i(\vec k_L\cdot\vec x-\omega_Lt)}\psi_0+(L\leftrightarrow R)\psi_1]
\ee
where $H_0=H_l+g_{00}|\psi_0|^2+g_{01}|\psi_1|^2$,
$H_1=H_l+g_{11}|\psi_1|^2+g_{10}|\psi_0|^2$ and 
$g_{ij}=4\pi\hbar^2a_{ij}/m$. $a_{ij}\sim5.5nm$~\cite{Savage} 
are the s-wave scattering lengths between atoms in $i$ and $j$ levels. 
We have called
$H_l=-(\hbar^2/2m)\nabla^2+V$ 
the one atom term of the Hamiltonian, where $V$ is the trapping potential.
The field amplitudes $E_{L,R}$ are rescaled in order to be expressed in
frequency units. They are thus the Rabi frequencies of the one-photon 
transition.
$\hbar\omega_2$ is the energy of the upper state of the one photon 
transition; $\hbar\omega_{0,1}$ are the energies of
the $m_F=-1,1$ levels, respectively. We set $\omega_0=0$.
In the adiabatic approximation, $\psi_2$ can be expressed in terms of 
$\psi_1$ and $\psi_0$ as 
$\psi_2=-[E_L^*e^{i(\vec k_L\cdot\vec x-\omega_Lt)}\psi_0+
E_R^*e^{i(\vec k_R\cdot\vec x-\omega_Rt)}\psi_1]/\Delta$,
where $\Delta=\omega_2-\omega_L$.
Thus we have two closed equations for $\psi_0$ and $\psi_1$. 

We introduce the gauge transformation
\be\label{trans1}
\tilde\psi_{0,1}=e^{-i\int\frac{|E_L|^2+|E_R|^2}{\Delta}dt}
e^{\pm i(\frac{\vec k_d}{2}\cdot\vec x-\frac{\omega_d}{2}t\pm\frac{\hbar\vec
k_d^2}{8m}t)}\psi_{0,1},
\ee
where $k_d=\vec k_L-\vec k_R$ and $\omega_d=\omega_L-\omega_R$.
As a result, the equations of motions become
\be\label{eq4}
i\hbar\dot{\tilde\psi}_{0,1}=\left(H_{0,1}\mp\frac{\hbar\delta}{2}
\right)\tilde\psi_{0,1}-\hbar\Omega\tilde\psi_{1,0}\pm\frac{i\hbar^2\vec k_d\cdot\vec
\nabla}{2m}\tilde\psi_{0,1}
\ee
Here, $\Omega\equiv\frac{E_LE_R^*}{\Delta}$ is the two photon Rabi frequency,
taken for simplicity as time independent and real, and the frequency $\delta$
is given by 
$\delta=\omega_1-\omega_d+(|E_L|^2-|E_R|^2)/\Delta$.
We assume $|E_L|^2=|E_R|^2$ and $\omega_d=\omega_1$, hence $\delta=0$.

If the number of atoms is sufficiently small, we can neglect the nonlinear
terms. Furthermore, let us initially consider a spatially homogeneous
condensate (no trap potential).
As a consequence, Eqs.~(\ref{eq4}) reduce to two linear
equations with constant coefficients, and the eigenvalue problem in the
reciprocal space is ruled by two linear algebraic equations for the transformed
fields $\phi_0(\vec k)$ and $\phi_1(\vec k)$. The momenta of the atoms in 
the two levels $m_F=-1$ and $m_F=1$ are respectively $\hbar(\vec k-\vec k_d/2)$ and 
$\hbar(\vec k+\vec k_d/2)$. 

\begin{figure}
\epsfig{width=5.5cm,figure=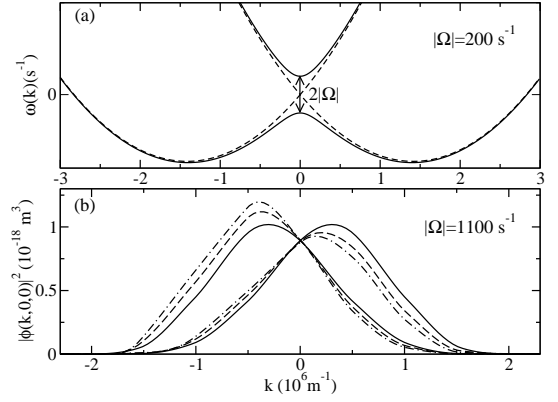,angle=-90}
\caption[]{(a) Dispersion curves of a free atom for
$|\vec k_d|=2\pi\cdot4.5\cdot10^5m^{-1}$ and $\Omega=200s^{-1}$.
(b) Density distributions $|\phi_{0,1}|^2$ for $k_d=2\pi\times4.5\cdot10^5 m^{-1}$,
$\Omega=1100 s^{-1}$, $\omega_\|=\omega_\bot=2\pi\cdot100 s^{-1}$
and three different boson numbers: $N=1390$ (solid), $1420$ (dashed), $1460$
(dot-dashed).}
\label{fig1}
\end{figure}

Solving the eigenvalue problem we find the two dispersion curves
$\hbar\omega(\vec k)=\hbar^2\vec k^2/(2m)\pm\{[\hbar^2
\vec k_d\vec k/(2m)+\hbar\delta/2]^2+\hbar^2|\Omega|^2\}^{1/2}$,
plotted in Fig.~\ref{fig1}a for $\Omega=200s^{-1}$ and $|\vec k_d|=2\pi\cdot
4.5\cdot10^5m^{-1}$.
As shown in the figure, the electromagnetic coupling modifies the parabolic
dispersion curves associated with the two hyperfine levels, lifting the
degeneracy at their intersection point. The energy gap for $\delta=0$ is
$2\hbar|\Omega|$. By varying $\delta$ one can rise or lower the energy 
separation between the two minima. For $\delta=0$ the two minima have the same 
energy. Introducing the harmonic trap potential, the ground state has no longer
a definite momentum. Furthermore, the two wells of the  dispersion 
curve are equally populated by quantum tunneling (Fig.~\ref{fig1}b, solid line).

%\section{Symmetry breaking.}

When the number of atoms is sufficiently high, a symmetry breaking occurs because
of the atomic interactions. We report in Fig.~\ref{fig1}b the distributions $|\phi_0|^2$ and 
$|\phi_1|^2$ for $\delta=0$ and for three different values of the number of atoms $N$. 
$\phi_{0,1}$ are the Fourier transforms of the ground state solution 
$(\tilde\psi_0,\tilde\psi_1)$ of Eqs.~(\ref{eq4}) for a spherical trap potential $V$
corresponding to equal longitudinal ($\omega_\bot$) and radial ($\omega_\|$) trap 
frequencies.
As it results, this interaction clusters the majority of atoms within a single well,
thus contrasting the quantum tunneling across the barrier. 
Due to the geometry of the problem, and taking into account that the 
scattering lengths $a_{ij}$ are practically equal, there is another ground 
state which is obtained from that of Fig.~\ref{fig1}b by inverting the horizontal 
axis and interchanging $\phi_0$ and $\phi_1$. Thus we have two stable 
stationary states with equal energy. 
The numerical evidence of Fig.~\ref{fig1}b is also supported by a synthetic 
variational argument, already exploited in Ref.~\onlinecite{OnlyMine} for $Li$, 
and based upon a suitable two mode approximation.

The matter field fluctuations allow the passage from one to the other state. If the 
decoherence is negligible, coherent oscillations between such states can be observed,
demonstrating MQC.
To evaluate the oscillation frequency we quantize the two mode system, as done
in Ref.~\onlinecite{OnlyMine}.
First of all, we write the classical Hamiltonian corresponding to the
equations of motions (\ref{eq4}), 
taking $g_{00}=g_{11}=g_{01}\equiv g$
\bey\label{hami}
\nonumber 
{\cal H}=\int\bigg[\psi_0^*H_l\psi_0+\psi_1^*H_l\psi_1+
\frac{g}{2}(|\psi_0|^2+|\psi_1|^2)^2+ \\
\frac{i\hbar^2\vec k_d}{2m}(\psi_0^*\vec\nabla\psi_0-\psi_1^*\vec\nabla
\psi_1)-\hbar\Omega(\psi_1^*\psi_0+\psi_0^*\psi_1)\bigg]d^3x
\eey

(from now on we omit the tilde on the $\psi$'s, even though we are
always in the gauge Eq.~(\ref{trans1})).
We then introduce the spinorial ground states
\be
\vec\psi_{g,1}(\vec x)\equiv\left(
\begin{array} {c}
                      \psi_{0g}(\vec x) \\
                      \psi_{1g}(\vec x)
\end{array}\right);
\vec\psi_{g,2}(\vec x)\equiv\left(
\begin{array} {c}
                      \psi_{1g}(-\vec x) \\
                      \psi_{0g}(-\vec x)
\end{array}
\right)
\ee
where $\psi_{0g}$ and $\psi_{1g}$ are the ground state wave-functions
associated with the two internal states. $\vec\psi_{g,2}$ is obtained from
$\vec\psi_{g,1}$ by interchanging the spinorial components and inverting the axes.

\begin{figure}
\epsfig{figure=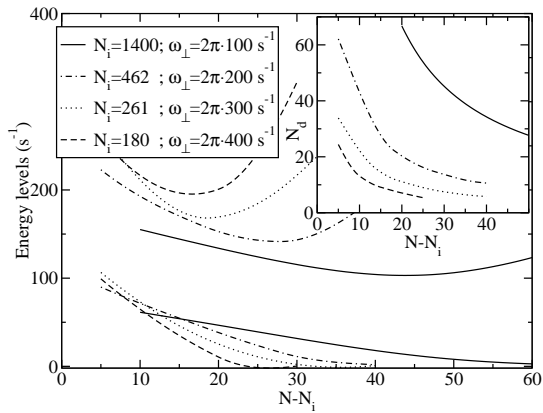,width=5.5cm,angle=-90}
\caption{First and second excited energy levels of the rubidium BEC versus the excess of atoms
above the breakup value $N_i$ for $\omega_\|=2\pi\cdot100s^{-1}$ and some values of
$\omega_\bot$. In the inset we plot $N_d$ for the same values of the trap frequencies.}
\label{fig2}
\end{figure}

\begin{figure}
\epsfig{figure=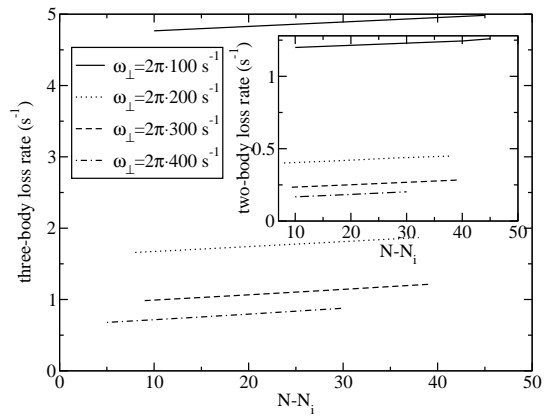,width=5.5cm,angle=-90}
\caption{Three-body and two-body (inset) loss rates. The latter one refers to atoms 
trapped in the $F=1,m=-1$ level and represents an upper limit~\cite{cornell}.}
\label{fig3}
\end{figure}

It is convenient to use the basis vectors $\vec\psi_a=\vec\psi_{g,1}+\vec\psi_{g,2}$
and $\vec\psi_b=\vec\psi_{g,1}-\vec\psi_{g,2}$.
We write the quantized spinorial field of the bicondensate as
$\vec\psi(\vec x)=\hat a\vec\psi_a(\vec x)+\hat b\vec\psi_b(\vec x)$,
and substitute this expression in the operator version of Eq.~(\ref{hami}). 
Thus, we obtain a reduced Hamiltonian ${\cal H}_{red}$ of the same
form found in the attractive case of Ref.~\onlinecite{OnlyMine}.
By ${\cal H}_{red}$, we evaluate the 
difference between the lowest eigenvalues, which provides the tunneling rate.
In Fig.~\ref{fig2} we plot the first and second energy level as a function of the
number of atoms and for some values of $\omega_\bot$. The energy of the ground state
is set to zero, so the energy of the first level gives the tunneling frequency. 
The entanglement between the condensate and the lost atoms induces a decoherence of
the superposition. Using the approach of Ref.~\onlinecite{OnlyMine}, we find that the 
coherence is given by $\tilde C=e^{-\epsilon M}$, where $M$ is the number of lost
atoms and $\epsilon=1-2\int d^3x\psi_0^*(\vec x)\psi_1(-\vec x)/N$.
The quantity $N_d=1/\epsilon$ is the number of atoms which must be lost
in order to reduce the coherence by $1/e$.
The inset of Fig.~\ref{fig2} shows how $N_d$ scales with
$N$. The relevant loss processes are two-body inelastic and three-body collisional 
decays. In Fig.~\ref{fig3} we report the average three-body and two-body (inset) loss 
rates, the latter one refers to atoms in the $m=-1$ level. 
We have used the upper limit of $1.6\cdot10^{-16}cm^3/s$ for the two-body loss rate 
coefficient and $5.8\cdot10^{-30}cm^6/s$ for the three-body processes, both 
of them measured in Ref.~\onlinecite{cornell} for the trapped Zeeman level $F=1,m=-1$.
Two-body decay can occur also by means of collisions between atoms in different Zeeman
levels. We suppose that the corresponding loss rate is of the same order of 
magnitude as the measured value. 
From Figs.~\ref{fig2},\ref{fig3} we find that the decoherence 
effects are negligible during a MQC oscillation period. For high $\omega_\bot$ the 
threshold $N_i$ 
and the loss rates are lower. For $\omega_\bot=2\pi\cdot200\div400s^{-1}$ the overall
loss rate is much smaller than the
corresponding tunneling frequencies. Therefore we can observe many oscillations before
a single atom is lost. For $\omega_\bot=2\pi\cdot100s^{-1}$ the loss rate is $>5 s^{-1}$. With
a tunneling frequency of $8 Hz$ also in this case we can observe an oscillation before
a single atom is lost. If the loss of atoms does not transfer energy to the trapped atoms,
the escape of a few atoms does not reduce the superposition coherence (inset of 
Fig.\ref{fig2}), but modifies slightly 
the tunneling rate. In Ref.~\onlinecite{guery} it is shown that inelastic collisional
processes induce local variations of the mean-field interparticle interaction and are
accompanied by the creation/annihilation of elementary excitations. This phenomenon
depends on the density and is completely negligible in our case.
Notice that Eqs.~(\ref{eq4}) are invariant if $N$ is varied by a factor $\alpha$ 
and the lengths and the energies are multiplied by $\alpha$ and $\alpha^{-2}$, 
respectively. So for $\omega=\omega_\bot=2\pi\cdot70s^{-1}$ the threshold is $N_i=1980$.
By the two mode model we find that the tunneling frequencies are reduced by nearly 
a factor $0.5$. However the two-body and three-body decay rates are reduced by 
the factors $0.25$ and $0.125$, respectively.

\begin{figure}
\epsfig{figure=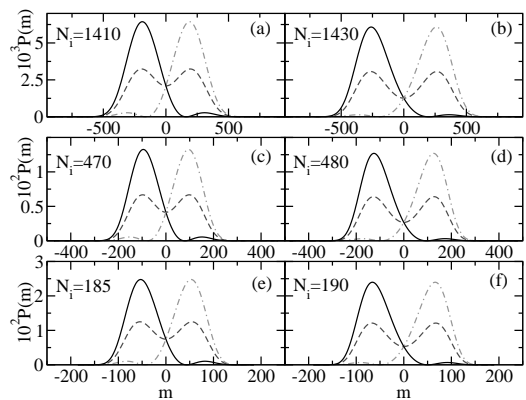,width=5.5cm,angle=-90}
\caption{Distribution $P(m)$ for some parameters. The solid, dashed and dashed-dot lines 
correspond to $P(m)$ at the initial time, at a quarter of period and half a period,
respectively. $\omega_\bot$ is (a-b) $2\pi\cdot100s^{-1}$,(c-d) $2\pi\cdot200s^{-1}$ and
(e-f) $2\pi\cdot400s^{-1}$.}
\label{fig4}
\end{figure}

As we have shown in Ref.~\onlinecite{OnlyMine}, the probability distribution $P(m)$ associated with
the observable $\hat M=\hat a^\dagger\hat b+\hat a\hat b^\dagger$ displays two peaks for
the ground state, that are the alive and dead state of the Schr\"odinger cat.
This observable is associated with the population measurement in one of the two
Zeeman levels. To observe the coherence we have to put the system in one of the two 
states, that corresponds to
take a superposition of the ground and first excited state of the condensate.
This can be obtained measuring with a nondemolition technique (e.g. a phase contrast
technique) the population in a Zeeman level when only the ground
state is populated. This observation collapses the $P(m)$ distribution to one peak,
as discussed in Ref.~\onlinecite{OnlyMine}. 
If the energy transfer is not too large we expect that only the first two energy 
states are populated. With this initial preparation the system begins to oscillate 
at the frequencies of Fig.~\ref{fig2} between the two Schr\"odinger cat states, 
as reported in Fig.~\ref{fig4}. At the initial time only one peak is present (solid line).
At a quarter of period corresponding to the frequency separation between ground and
first excited state, the $P(m)$ displays two peaks (dashed line). At half a period
the only peak is that absent at the initial time, thus there is a coherent
oscillation between the two states. Detecting such an oscillation would provide
evidence of a SC at an intermediate time when both peaks are present.

Notice that if we had $N$ loosely coupled or independent atoms
($Ng$ sufficiently low or even zero) the superposition of ground
and first excited state would have a single peak, oscillating
as a coherent state inside a harmonic potential. This would
by no means be a SC. On the contrary, we have shown that, for
$Ng$ sufficiently high, we have a two peak distribution with
the two partial barycenters at nearly constant positions.
During the evolution, the two peak amplitudes oscillate, that
is, the probability to find the system in either state
oscillates.     

In conclusion, we have shown that a double well dispersion curve can be obtained
by a suitable Raman coupling. In this situation a symmetry breaking in momentum
space is demonstrated solving two coupled Gross-Pitaevskii equations. The condensate 
can oscillate between the two emerging asymmetrical steady solutions (SC states) 
by means of the field quantum fluctuations (MQC).
We have found that it is possible to obtain an oscillation
frequency between the SC states around $50\div100 s^{-1}$. In order to neglect
the thermal activation, the second excited level energy ($E_2$) has to be higher than the
thermal energy. From Fig.~\ref{fig2} we can see that $E_2$ ranges between
$100\div300 s^{-1}$, that correspond to a temperature of $0.7\div2.3 nK$. 
If the cooling is performed below the threshold, when the symmetry breaking does not
occur, the required temperature can be $\sim5 nK$. 
However it may not be
necessary to cool at very low temperatures the whole condensate, but just the 
involved degrees of freedom, provided that this latter ones are 
weakly coupled with the other modes, which act as a thermal bath. 
We remark that for low densities the evaporative cooling allows to reach much lower
temperatures, because of a smaller three-body decay rate. 
To be sure that no excitation is present one could tailor the trapping potential
in such a way that only the first two levels are bound.
In this work we have chosen the parameters for which the tunneling frequency is
much larger than the decoherence rate, however the symmetry breaking and the 
super-Poissonian atom fluctuations below threshold can be observed with a much
higher number of atoms, thus these phenomena are observable with the present 
technology.

\end{document}